\documentclass[a4paper]{article}
\usepackage{amsmath,amssymb,amsfonts,bm,xcolor}

\usepackage{algorithm,algorithmic,url}
\newcommand{\algrule}[1][.2pt]{\par\vskip.5\baselineskip\hrule height #1\par\vskip.5\baselineskip}
\newcommand{\amax}[2]{\underset{#1}{\mathrm{arg}\,\mathrm{max}}\,\,#2}
\newcommand{\dd}{\, \textrm{d}}       
\newcommand{\dint}{\int \!}       

\usepackage{enumitem,hyperref,natbib,multirow,rotating,sectsty,booktabs}
\allsectionsfont{\sffamily}
\usepackage[margin=1in]{geometry}

\usepackage{setspace}

\title{Constructing Metropolis-Hastings proposals\\ using damped BFGS updates}

\author{Johan Dahlin, Adrian~Wills and Brett~Ninness%
\thanks{E-mail adresses to authors: \url{firstname.lastname@newcastle.edu.au}. JD and AW are with the School of Engineering, The University of Newcastle, Australia. BN is with the Faculty of Engineering and Built Environment, The
University of Newcastle, Australia. This work was supported by the Australian Research Council Discovery Project DP140104350.}%
}

\begin{document}
\maketitle

\doublespacing

\begin{abstract}
	\noindent
	The computation of Bayesian estimates of
	system parameters and functions of them on the basis of observed
	system performance data is a common problem within system identification.
	This is a previously studied issue where
	stochastic simulation approaches have been examined using the
	popular Metropolis--Hastings (MH) algorithm.  This prior study has
	identified a recognised difficulty of tuning the \emph{proposal
	distribution} so that the MH method provides realisations with
	sufficient mixing to deliver efficient convergence.  This paper
	proposes and empirically examines a method of tuning the proposal
	using ideas borrowed from the numerical optimisation literature
	around efficient computation of Hessians so that gradient and
	curvature information of the target posterior can be incorporated in
	the proposal. \\

\end{abstract}

\newpage
\maketitle

\section{Introduction}
\label{sec:intro}
State-space models (SSMs) are ubiquitous within many scientific disciplines including system identification \citep{Ljung1999} and finance \citep{DurbinKoopman2012}.
A common problem within SSMs is the estimation of the parameters $\theta \in \mathbb{R}^{n_{\theta}}$ given some observed data $y_{1:T} = \{y_1,\cdots,y_T\}$.
In this paper, we consider this problem for SSMs expressed by%
\begin{align}
	x_0
	\sim
	\mu_{\theta}(x_0),
	\quad
	x_{t+1} | x_t
	\sim
	f_{\theta}(x_{t+1} | x_{t}),
	\quad
	y_{t} | x_t
	\sim
	g_{\theta}(y_{t} | x_{t}),
\label{eq:ssm:definition}%
\end{align}%
where $\mu_{\theta}$, $f_{\theta}$ and $g_{\theta}$ denote known densities parameterized by $\theta$.
Here, $x_t \in \mathbb{R}^{n_x}$ and $y_t \in \mathbb{R}^{n_u}$ denote the state and the observation from the system at time $t$.
Note that this parameterisation \eqref{eq:ssm:definition} includes most non-linear and non-Gaussian SSMs and an input $u_t \in \mathbb{R}^{n_u}$ can be added as arguments to $f_{\theta}$ and $g_{\theta}$.

One common approach for estimating $\theta$ in \eqref{eq:ssm:definition} is to employ the implied one-step-ahead prediction distribution $p_{\theta}(y_t| y_{t-1})$ to form the likelihood
\begin{align}
	\pi(\theta)
	=
	p_{\theta}(y_{1:T})
	=
	p_{\theta}(y_1)
	\prod_{t=2}^T
	p_{\theta}(y_t| y_{1:t-1}),
	\label{eq:likelihood:def}
\end{align}
of the observed data.
Then an estimate of $\theta$ is given by the maximising argument of the likelihood, i.e.,
\begin{align}
	\widehat{\theta}_{\text{ML}}
	=
	\amax{\theta}{p_{\theta}(y_{1:T})},
	\label{eq:maximumlikelihood:est}
\end{align}
which is the well-known maximum likelihood (ML) estimate~\citep{Ljung1999}.
In practice, $p_{\theta}(y_t| y_{t-1})$ is intractable for most SSMs but can be estimated unbiasedly using particle methods~\citep{DoucetJohansen2011},
These methods can be employed within e.g., gradient-based optimisation~\citep{PoyiadjisDoucetSingh2011} and the Expectation--Maximisation algorithm~\citep{SchonWillsNinness2011} to approximately solve the non-convex intractable problem in~\eqref{eq:maximumlikelihood:est}.

In this paper, we take another approach to estimate $\theta$ by using the Bayesian paradigm~\citep{Peterka1981,Robert2007}.
This amounts to computing the posterior
\begin{align}
	p(\theta|y_{1:T})
	\propto
	p_{\theta}(y_{1:T})\,
	p(\theta),
	\label{eq:posterior}
\end{align}
where $p(\theta)$ is a prior distribution for $\theta$ that can be used to encode prior user information about the parameters.
However, this posterior is intractable as the likelihood \eqref{eq:likelihood:def} cannot be computed in closed-form.

Instead, we make use of a stochastic simulation approach to address this difficulty which constructs a random number generator such that
\begin{align}
	\theta_k
	\sim
	\pi(\theta).
	\label{eq:posterior:sampling}
\end{align}
This is a widely applied approach in the statistics literature where there has been an explosion of its use in applications in the last decade.

A standard approach to building a suitable random number generator is to employ (particle) Metropolis--Hastings (MH; \citealp{RobertCasella2004,AndrieuDoucetHolenstein2010}) which is a very general algorithm for computing realisations from $\pi(\theta)$ if it can be evaluated point-wise.
Unfortunately achieving reasonable convergence requires careful tuning of the algorithm, which essentially entails leveraging information about the unknown posterior.

This difficulty is well recognised within the literature which is usually mitigated by employing adaptive methods~\citep{AndrieuThoms2008} and the inclusion of geometric information~\citep{GirolamiCalderhead2011}.
The latter approach requires the computation of the Hessian of the log-posterior, which can be challenging to compute directly even for a linear Gaussian SSMs where standard Kalman methods are applicable.

The problem is even worse when particle methods \citep{DoucetJohansen2011} are employed to estimate the latent state and the likelihood.
This is the result of the empirical observation that the Hessian estimates obtained using particle methods often are noisy and inaccurate even using a large amount of particles.
However, sometimes the gradient estimates are accurate even using a small amount of particles.
It is therefore of interest to study the problem of estimating the Hessian using noisy gradient information.

The contribution of this paper is explore the use of damped \emph{Brodyen--Fletcher--Goldfarb--Shanno} (BFGS) updates for approximating the local curvature encoded by the Hessian using only gradient information.
Ths is very widely employed approach for solving smooth numerical optimisation problems~\citep{NocedalWright2006}.

Related work regarding the use of BFGS within MH includes \cite{ZhangSutton2011} where the authors apply this idea to regression problems.
This work is extended to a class of SSMs with intractable likelihoods by~\cite{DahlinLindstenSchon2015b}.
The major novelty in the present paper comes from using the damped BFGS update to ensure that the Hessian is positive semi-definite even when particle methods are employed.
Finally, the proposed approach delivers superior performance compared to earlier attempts to make use of BFGS within MH.
Hence, we obtain good performance of MH without the need for tedious user-tuning which is a step towards automated Bayesian inference methods.


\section{Sampling from the posterior}
\label{sec:overview}
There exists a suite of so-called \emph{Markov chain Monte Carlo} (MCMC; \citealp{RobertCasella2004}) methods for constructing a Markov chain producing realisations $\{\theta_k\}$ with user-specified invariant distribution $\pi(\theta)$.
Since under mild assumptions realisations from Markov chains have distribution convergent to the invariant distribution of the chain, this provides a means to build a random number generator~\eqref{eq:posterior:sampling} with arbitrary target distribution $\pi$.

These samples can then be employed to approximate the posterior.
Given the posterior, point estimates such as the conditional mean
\begin{align}
	\widehat\theta_{\text{CM}}
	=
	\mathbb{E} \big[ \theta \big| y_{1:T} \big]
	=
	\dint
	\theta
	\,
	\pi(\theta)
	\,
	\dd \theta,
	\label{eq:bayes:conditional:mean}
\end{align}
can be obtained.
These are of interest as they possess a minimum mean squared error property and does not rely on asymptotic results as the ML estimator~\eqref{eq:maximumlikelihood:est}.
Moreover, error bounds for each element $i$ of the estimated parameter vector can be obtained by computing the (marginal) posterior density
\begin{align}
	p(\theta^{i}| y_{1:T})
	=
	\dint
	\pi(\theta)
	\,
	\dd \theta^{-i},
	\label{eq:bayes:marginal:posterior}
\end{align}
where $\theta^{-i}$ denotes the vector $\theta$ without its $i$th element.
Unfortunately, both \eqref{eq:bayes:conditional:mean} and \eqref{eq:bayes:marginal:posterior} require the evaluation of multidimensional integrals, which can be computationally challenging, particularly as $n_{\theta}=\mbox{dim}\{\theta\}$ grows.

This results in that the expected value of any arbitrary (measurable) function $\varphi: \mathbb{R}^{n_{\theta}} \rightarrow \mathbb{R}$ given by
\begin{align*}
	\pi[\varphi]
	=
	\mathbb{E} \left[ \varphi(\theta) \right]
	=
	\dint
	\varphi(\theta)
	\,
	\pi(\theta)
	\,
	\dd \theta,
\end{align*}
can be approximated by
\begin{align}
	\widehat{\pi}^K[\varphi]
	=
	\frac{1}{K}
	\sum_{k=1}^K
	\,
	\varphi(\theta_k),
	\label{eq:empirical:app}
\end{align}
using samples from the random number generator.
Furthermore, we have that the estimator obeys the strong law of large numbers and is consistent, i.e.,
\begin{align}
	\widehat{\pi}^K[\varphi]
	\stackrel{\text{a.s.}}{\longrightarrow}
	\pi[\varphi],
	\quad
	K \rightarrow \infty.
	\label{eq:slln}
\end{align}
Choosing $\varphi(\theta)=\theta$ then gives an approximation to the conditional mean estimate (\ref{eq:bayes:conditional:mean}) and choosing $\varphi(\theta)$ as an appropriate indicator function then gives approximations of $\pi(\theta)$ as sample histograms.

The MH algorithm is arguably one of the most widely employed MCMC techniques to implement \eqref{eq:posterior:sampling}.
It operates by taking an arbitrary \emph{proposal} Markov chain $q(\theta_k| \theta_{k-1})$ and modulating it by randomly accepting realisations from this \emph{base} chain.
The acceptance probability depends on the target $\pi(\theta)$.
In theory (but not in practice) the proposal distribution can be selected quite freely but in the majority of cases a Gaussian proposal is used,
\begin{align}
	q \big( \theta' | \theta_{k-1} \big)
	&=
	\mathcal{N}
	\Big(
	\theta';
	\mu \big( \theta_{k-1} \big),
	\Sigma \big( \theta_{k-1} \big)
	\Big),
	\label{eq:mh:proposal}
\end{align}
where $\theta'$ denotes the candidate parameter with $\mu$ and $\Sigma$ denoting a mean and covariance function, respectively.

After generating a candidate parameter via \eqref{eq:mh:proposal}, it
is accepted with the acceptance probability
\begin{align}
	\alpha(\theta', \theta_{k-1})
	=
	1
	\wedge
	\frac{\pi(\theta')}{\pi(\theta_{k-1})}
	\frac{q(\theta_{k-1} | \theta')}{q(\theta' | \theta_{k-1})},
	\label{eq:mh:acceptprob}
\end{align}
where $a \wedge b = \min(a,b)$.
We set $\theta_k \leftarrow \theta'$ if the candidate parameter is accepted and $\theta_k \leftarrow \theta_{k-1}$ if it is rejected.
Note that, we are only required to be able to point-wise evaluate $\pi(\theta)$ to implement MH as presented in Algorithm~\ref{alg:mh}.

\begin{algorithm}[!t]
	\caption{\textsf{Metropolis-Hastings (MH)}}
	\textsc{Inputs:} $K>0$, $\theta_0$ and $q$. \textsc{Output:} $\{\theta_1,\ldots,\theta_k\}$.
	\algrule[.4pt]
	\begin{algorithmic}[1]
		\STATE Compute $\pi(\theta_0)$.
		\FOR{$k=1$ to $K$}
			\STATE Sample $\theta' \sim q(\theta' | \theta_{k-1})$ using \eqref{eq:mh:proposal}.
			\STATE Compute $\pi(\theta')$ using Kalman or particle methods.
			\STATE Sample $\omega_k$ uniformly over $[0,1]$.
			\IF{$\omega_k \leq \min\{1,\alpha(\theta',\theta_{k-1})\}$ given by \eqref{eq:mh:acceptprob}.}
				\STATE Accept $\theta'$, i.e.\
				$\theta_k \leftarrow \theta'$.
			\ELSE
				\STATE
				Reject $\theta'$, i.e.\
				$\theta_k \leftarrow \theta_{k-1}$.
			\ENDIF
		\ENDFOR
	\end{algorithmic}
	\label{alg:mh}
\end{algorithm}

An essential point is that the speed of the convergence (\ref{eq:slln}) depends on how correlated the realisations $\{\varphi(\theta_k)\}$ are.
The more uncorrelated the faster the convergence and hence the better the approximation (\ref{eq:empirical:app}) for a given finite number $K$ of realisations.
This is well understood in the MCMC literature, where the variance in the stochastic approximation (\ref{eq:empirical:app}) is established as being proportional to the \emph{integrated autocorrelation}
\begin{align}
\textsf{IF}
=
1
+
2
\sum_{k=2}^{\infty}
\textsf{corr}
\{
	\varphi(\theta_1),
	\varphi(\theta_k)
\},
\label{eq:if}
\end{align}
which also is known as the inefficiency factor (IF).
In turn, the correlation of the realisations $\{\varphi(\theta_k)\}$ is critically dependent on the choice of the proposal $q(\theta^{\prime}|\theta_{k-1})$.

From the work of~\cite{GirolamiCalderhead2011} it is known that the \textit{mixing} (i.e.\ autocorrelation) can be greatly improved by the inclusion of gradient and curvature information regarding the posterior into the proposal~\eqref{eq:mh:proposal}.
This is especially important when the posterior is non-isotropic, i.e., some parameters influence the value of the posterior to a larger degree than others.
The influence can also vary over the parameter space, which makes local information about the curvature important to increase mixing.


\section{Hessian estimation}
A significant further problem is that the curvature information is difficult to
obtain in an efficient manner for general SSMs when particle methods are employed within MH as discussed in the introduction.
The gradient of the log-posterior is quite simple to estimate efficiently using \textit{Fisher's identity} \citep{CappeMoulinesRyden2005},
\begin{align}
	G(\theta)
	&=
	\nabla
	\log
	p_{\theta}(y_{1:T})
	=
	\nabla\,
	\mathbb{E}_{\theta}
	\left[
		\log p_{\theta}(x_{1:T}, y_{1:T})
	\Big|
	y_{1:T}
	\right],
\label{eq:FishersIdentity}%
\end{align}
\noindent where either a Kalman or particle smoother can be employed to compute or
approximate the expectation.

The \textit{Louis' identity}~\citep{CappeMoulinesRyden2005} can be used in an analogue manner to estimate the negative Hessian of the log-posterior .
Unfortunately, the resulting estimator often suffers from large noise sensitivity which results in frequent loss of positive definiteness.
This is problematic as the inverse of the Hessian is often included in the proposal~\eqref{eq:mh:proposal} as its covariance.

Another approach is to compute an estimate of the Hessian by using pilot runs.
This results in a so-called pre-conditioning matrix, which can be used to scale the proposal.
However, results from numerical studies in e.g., \cite{GirolamiCalderhead2011}, \cite{NemethSherlockFearnhead2016} and \cite{DahlinLindstenSchon2015a} indicate that this parameter-independent approach is sub-optimal in terms of mixing.

\subsection{Damped limited-memory BFGS}
To address these problems, we propose to leverage knowledge from the optimisation literature, where the curvature information often is estimated from gradient information.
This is the approach used in the highly successful quasi-Newton algorithms that allow for optimising non-linear functions.
Again, this is of interest as gradient estimates often are accurate and relatively computationally cheap to obtain using particle methods in comparison to Hessian estimates of the posterior.

In classical mathematical optimisation, the class of so-called quasi-Newton methods~\citep{NocedalWright2006} were developed to incorporate curvature information into the search direction calculation in order to
accelerate convergence.
The details of such methods can be easily found in standard references such as \cite{NocedalWright2006}, but in essence, these methods utilise gradient and iterate information in order to form estimates of the Hessian, or its inverse.

One of the most celebrated quasi-Newton methods is the BFGS approach, which employs a rank-2 update to the current Hessian estimate $H_l$ to form a better estimate $H_{l+1}$ via the recursion
\begin{align}
	H_{l+1}
	&=
	(I - \rho_l s_l z_l^{\top} )
	H_l
	(I - \rho_l z_l s_l^{\top} )
	+
	\rho_l
	z_l
	z_l^{\top},
	\label{eq:bfgs:update}
	\\
	\nonumber
	\rho_l &= (z_l^{\top} s_l)^{-1},
	\quad
	s_l = \theta_l - \theta_{l-1},
	\quad
	z_l = G(\theta_l) -
G(\theta_{l-1}).
\end{align}
It can be observed that the Hessian estimate will remain positive-definite if $\rho_l > 0$ for all iterations.
In an optimisation setting this condition on $\rho_l$ can be guaranteed by a a line-search algorithm that satisfies Wolfe conditions, see \cite[Chapter~8]{NocedalWright2006}.
Unfortunately, in the MH setting, this cannot be enforced since it would result in a Markov chain that converges to a single point.

To ameliorate this problem, here we employ the so-called damped BFGS method where $z_l$ is replaced by $r_l$ via
\begin{align*}
	r_l &= \beta_l z_l + (1-\beta_l) H_l s_l, \\
	\beta_l
	&=
	\begin{cases}
	1,
	&
	\text{if } s_l^{\top} z_l \geq 0.2 s_l^{\top} H_l s_l\\
	(0.8s_l^{\top} H_l s_l)/(s_l^{\top} H_l s_l - s_l^{\top} z_l),
	&
	\text{if } s_l^{\top} z_l < 0.2 s_l^{\top} H_l s_l
\end{cases},
\\
	H_{l+1}
		&=
	(I - \rho_l s_l r_l^{\top} )
	H_l
	(I - \rho_l r_l s_l^{\top} )
	+
	\rho_l
	r_l
	r_l^{\top}.
\end{align*}%
In addition to the damping term, we further employ a limited memory~\citep[Chapter~9]{NocedalWright2006} implementation of the damped-BFGS approach so that the computational load remains modest.


\section{Quasi-Newton-based proposals}
\label{sec:qnproposals}
\label{sec:qnproposals:algorithm}
To construct a good MH proposal, the gradient information together with the Hessian estimate from BFGS will be used in the mean and covariance function entering \eqref{eq:mh:proposal}.
A typical choice resulting from a second-order Taylor expansion \citep{DahlinLindstenSchon2015a} of the log-posterior is
\begin{align}
	\mu(\theta)
	=
	\theta
	+
	\frac{\epsilon^2}{2}
	H^{-1}(\theta)
	G(\theta),
	\quad
	\Sigma(\theta)
	=
	\epsilon^2
	H^{-1}(\theta),
	\label{eq:mh:proposal:secondorder}
\end{align}
where $\epsilon > 0$ denotes a step size specified by the user.
Hence, we can see the proposal as a local Gaussian approximation of
the posterior, which should allow for efficient sampling from it.
Another way of motivating \eqref{eq:mh:proposal:secondorder} is to see
it as a random walk on a Riemannian manifold, see
\cite{GirolamiCalderhead2011} for details.

The inclusion of the BFGS algorithm to estimate the Hessian requires us to make some changes to MH.
The main problem is that information from $M$ iterations back in the algorithm is used to construct the proposal distribution.
In the standard version of MH, only information from the last iteration is allowed to be used in the proposal due to the Markov property.

To solve this problem, we are required to extend the Markov chain from a first-order chain to an $M$-order chain.
This allows MH to retain its validity as discussed by \cite{ZhangSutton2011} and \cite{DahlinLindstenSchon2015b}.
The major algorithmic change to Algorithm~\ref{alg:mh} is that the gradient and Hessian is computed in Step~4 using a smoother and Algorithm~\ref{alg:qnproposal}.
Moreover, the proposal step in MH is replaced by sampling from
\begin{align}
	q \big( \theta' | \psi_{k,M} \big)
	&=
	\mathcal{N}
	\Big(
	\theta';
	\mu_{\text{QN}} \big( \theta_{k-M} \big),
	\epsilon^2 \Sigma_{\text{QN}}^{-1} \big( \psi_{k, M} \big)
	\Big),
	\label{eq:qmh:proposal}
	\\
	\mu_{\text{QN}}(\theta_{k-M})
	&=
	\theta_{k-M}
	+
	\frac{\epsilon^2}{2}
	\Sigma^{-1}_{\text{QN}}(\psi_{k,M})
	G(\theta_{k-M}).
	\nonumber
\end{align}
using the procedure in Algorithm~\ref{alg:qnproposal} with $\psi_{k,M} \triangleq \{\theta_i, G(\theta_i)\}_{i=k-M}^k$.
Finally, we change Step~9 in Algorithm~\ref{alg:mh} to $\theta_{k} \leftarrow \theta_{k-M}$ when the candidate parameter is rejected due to that the proposal now is centered aroung $\theta_{k-M}$.

\begin{algorithm}[!t]
\caption{\textsf{Quasi-Newton proposal}}
\textsc{Inputs:} $\psi_{k,M} \triangleq \{\theta_i, G(\theta_i)\}_{i=k-M}^k$ and $\delta > 0$. \textsc{Output:} $\theta'$.
\algrule[.4pt]
\begin{algorithmic}[1]
\STATE Extract the $\bar{M}$ unique elements from $\psi_{k,M}$ and sort them in ascending order (with respect to the log-target) to obtain $\bar{\psi}_{k,M}$.
\IF{ $\bar{M} \geq 2$ }
	\STATE Initialise the Hessian estimate $H_0$.
	\FOR{$l=1$ to $\bar{M}$}
		\STATE Calculate $s_l$ and $z_l$ based on the $l$th pair in $\bar{\psi}_{k,M}$.
		\STATE Carry out the update \eqref{eq:bfgs:update} to obtain $H_l$.
	\ENDFOR
	\STATE Set $\Sigma_{\text{QN}}(\psi_{k,M}) = -H_{\bar{M}}(\theta')$.
	\ELSE
		\STATE Set $\Sigma_{\text{QN}}(\psi_{k,M}) = \delta \mathbf{I}_p$.
	\ENDIF
\STATE Sample from \eqref{eq:qmh:proposal} to obtain $\theta'$.
\end{algorithmic}
\label{alg:qnproposal}
\end{algorithm}


\section{Numerical illustrations}
\label{sec:results}
We provide three numerical illustrations to gain understanding about the proposed algorithm and compare it to other alternatives in the literature.
The implementation details are summarised in Appendix~\ref{app:impdetails} and the source code can be downloaded as described in Section~\ref{sec:conclusions}.

\subsection{LGSS model with synthetic data}
\label{sec:results:lgss:kalman}
We begin by considering a linear Gaussian state-space (LGSS) model as it is possible to solve the state inference problem exactly using the Kalman smoother.
This enables us to compute the log-posterior and its gradients using exact recursions, which will give an indication about the optimal performance of
various MH proposals.
The model is given by
\begin{subequations}
\begin{align}
	x_{t+1} | x_t
	&\sim
	\mathcal{N}
	\Big(
	x_{t+1};
	\mu + \phi (x_t - \mu),
	\sigma_v^2
	\Big),
	\\
	y_{t}  | x_t
	&\sim
	\mathcal{N}
	\Big(
	y_t;
	x_t, 0.5^2
	\Big),
\end{align}%
\label{eq:lgssmodel}%
\end{subequations}%
\noindent with $\theta=\{\mu,\phi,\sigma_v\}$ and $\mu \in \mathbb{R}$, $\phi \in (-1,1)$ and $\sigma_v \in \mathbb{R}_{+}$.
A synthetic data set consisting of a realisation with $T=500$ observations is simulated from the model using the parameters $\{0.2,0.5,1.0\}$.

We make use of $25$ Monte Carlo simulations using the same data to compute the IF \eqref{eq:if}, and estimate the computational time for different proposals.
Table~\ref{tbl:results:lgss} summarises the median results from these simulations, which include the acceptance rate, the fraction of Hessian estimates that are corrected, the maximum of IF and the time required for each iteration and the time required to obtain one effective sample from the posterior.
For the IF, we also provide the IQR (the distance between the $25\%$ and $75\%$ quantiles).

\begin{table}[t]
	\begin{center}
	\begin{tabular}{lll|ccc|cc}
	\toprule
	& & & & & & \multicolumn{2}{c}{Time} \\
	\cmidrule(r){7-8}
	& Alg. & Reg. & Acc. & Cor. & max \textsf{IF} &  Iter. & Samp. \\
	\midrule
	\parbox[t]{2mm}{\multirow{7}{*}{\rotatebox[origin=c]{90}{Kalman}}}
	&pMH0  &-    & 0.12 & -    & $89 \pm 62$  & 0.62 & 55 \\
	&pMH1  &-    & 0.37 & -    & $113 \pm 53$ & 1.14 & 130 \\
	\cmidrule(r){2-8}
	&dBFGS &-    & 0.76 & -    & $\mathbf{24} \pm 7$ & 2.16 & $\mathbf{53}$ \\
	&iBFGS &flip & 0.52 & 0.95 & $42 \pm 17$  & 1.83 & 77 \\
	&iBFGS &reg  & 0.45 & 0.95 & $34 \pm 15$  & 1.77 & 61 \\
	&iBFGS &hyb  & 0.60 & 0.95 & $40 \pm 17$  & 1.84 & 76 \\
	&eBFGS &hyb  & 0.62 & 1.00 & $34 \pm 19$  & 1.89 & 64 \\
	\midrule
	\parbox[t]{2mm}{\multirow{4}{*}{\rotatebox[origin=c]{90}{Particles}}}
	&pMH0  &-    & 0.04 & -    & $745 \pm 177$ & 0.3 & 211 \\
	&pMH1  &-    & 0.25 & -    & $275 \pm 131$  & 0.3 & 103 \\
	\cmidrule(r){2-8}
	&dBFGS &-    & 0.32 & -    & $\mathbf{39} \pm 29$   & 0.3 & $\mathbf{14}$ \\
	&iBFGS &hyb  & 0.33 & 0.93 & $79 \pm 43$   & 0.3 & 27 \\
	\bottomrule \\
	\end{tabular}
	\end{center}
	\caption{Performance statistics as the median over $25$ Monte Carlo runs for different proposals in MH.}
	\label{tbl:results:lgss}
\end{table}

The BFGS-type proposals are benchmarked against pre-conditioned versions of MH0 and MH1 denoted pMH0 and pMH1, respectively.
In pMH1, we set $H^{-1}(\theta) = \mathcal{P}$, where $\mathcal{P}$ denotes an estimate of the posterior covariance computed using pilot runs.
In pMH0, we use the same approach as for pMH1 but also set $G(\theta)=0$.

\begin{figure}[p]
	\centering
	\includegraphics[width=0.75\textwidth]{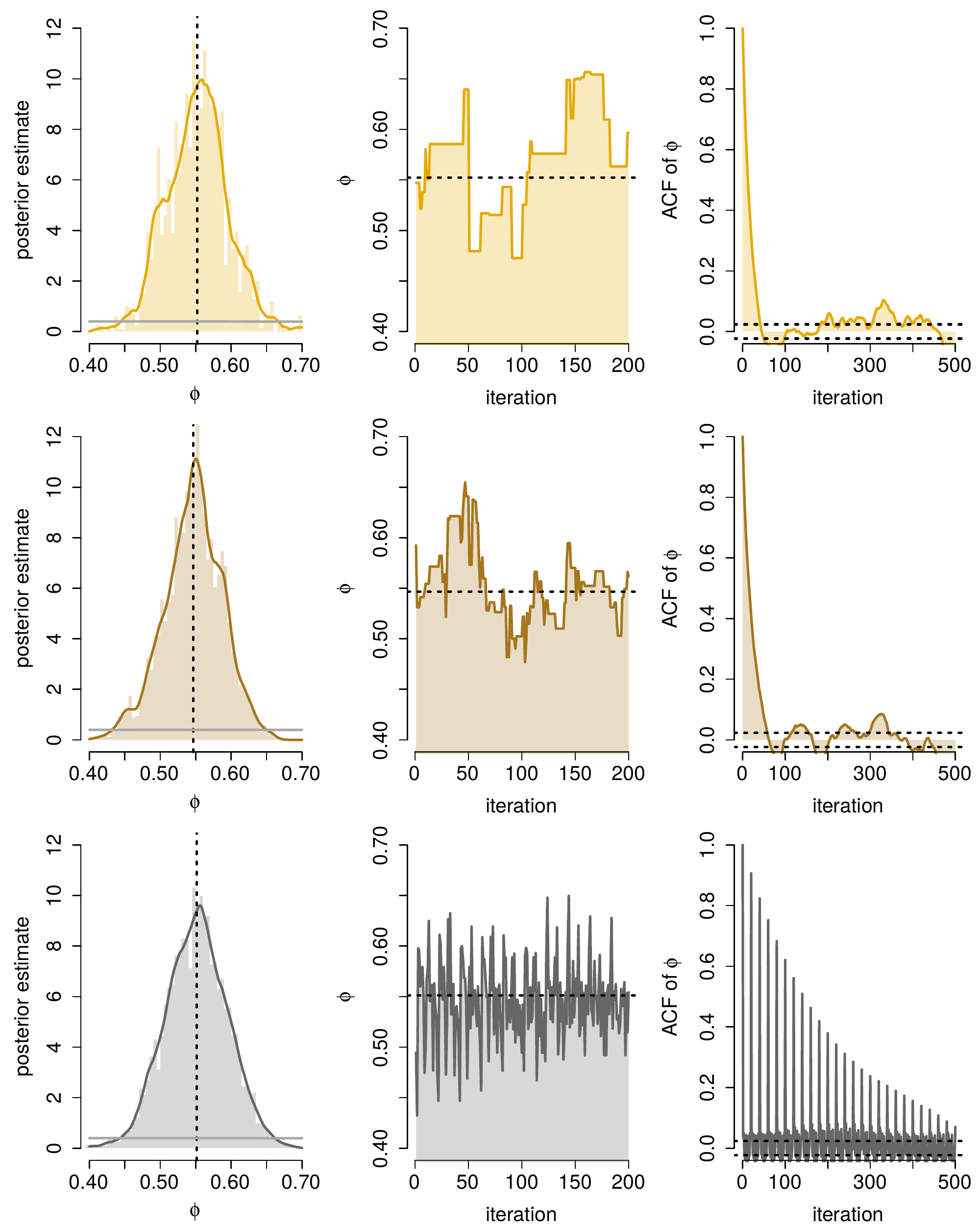}
	\caption{The posterior estimates (left), trace plot (center) and ACF estimate (right) for $\phi$ using pMH0 (top), pMH1 (middle) and dBFGS (bottom). The dashed lines in the left and center plots indicate the estimated posterior mean. The dashed lines in the right plot indicate the $95\%$ confidence intervals. The grey lines if the left plots indicate the prior distribution.}
	\label{fig:example1_lgss_kalman_paper}
\end{figure}

Three different BFGS proposals are used: (d)amped, (i)gnoring the curvature condition and (e)nforcing the curvature condition.
For the latter two, the Hessian estimates are often negative definite and therefore require some correction.
We apply the three different methods outlined in Appendix~\ref{app:hessian:corrections}: (flip)ping the negative eigenvalues, (reg)ularising the estimate and the (hyb)rid method.

The time per effective sample (in milliseconds) is presented in the right-most column in the table.
We note that the proposed method based on damped BFGS to approximate the curvature information locally performs the best.
It requires $53$ milliseconds to produce one sample from the posterior.
This is smaller than for pMH0, which would be the standard approach in this setting.
Furthermore, we note that the other BFGS-type proposals require a large amount of corrections of the Hessian, which is not desirable this might introduce numerical instability.

Figure~\ref{fig:example1_lgss_kalman_paper} presents the posterior estimate, Markov chain trace with its corresponding ACF for a particular case in the simulation study.
We note that the mixing is much better for the BFGS-based proposal (lower) compared with pMH0 (upper) and pMH1 (middle).
Furthermore, the ACF for the BFGS-based proposal exhibit a quite different behaviour compared with the other two proposals due to the $M$ step dependence in the Markov chain.
Comparing the posterior estimates, we conclude that the proposed method generates good estimates centered around the correct parameter and with reasonable variance.

\subsection{LGSS model revisited}
\label{sec:results:lgss:particles}
We repeat the same experiment when the Kalman filter and smoother is replaced by a particle filter and fixed-lag particle smoother as described in \cite{DahlinLindstenSchon2015a}.
Again, Table~\ref{tbl:results:lgss} summarises the results with all timings now
expressed in seconds.
The pMH0 and pMH1 perform worse in this case due to the noise in the estimates of the log-posterior and its gradients.

However, the proposed method performs well and generates one effective sample from the posterior every $14$ seconds.
This is a substantial decrease compared with the other methods and in particular an acceleration by at least a factor of two compared with the quasi-Newton approach proposed by \cite{ZhangSutton2011} and \cite{DahlinLindstenSchon2015b}.
Hence, the proposed method outperforms both the standard approaches to designing a proposal and with current state-of-the-art in quasi-Newton proposals.


\begin{figure}[p]
	\centering
	\includegraphics[width=0.75\textwidth]{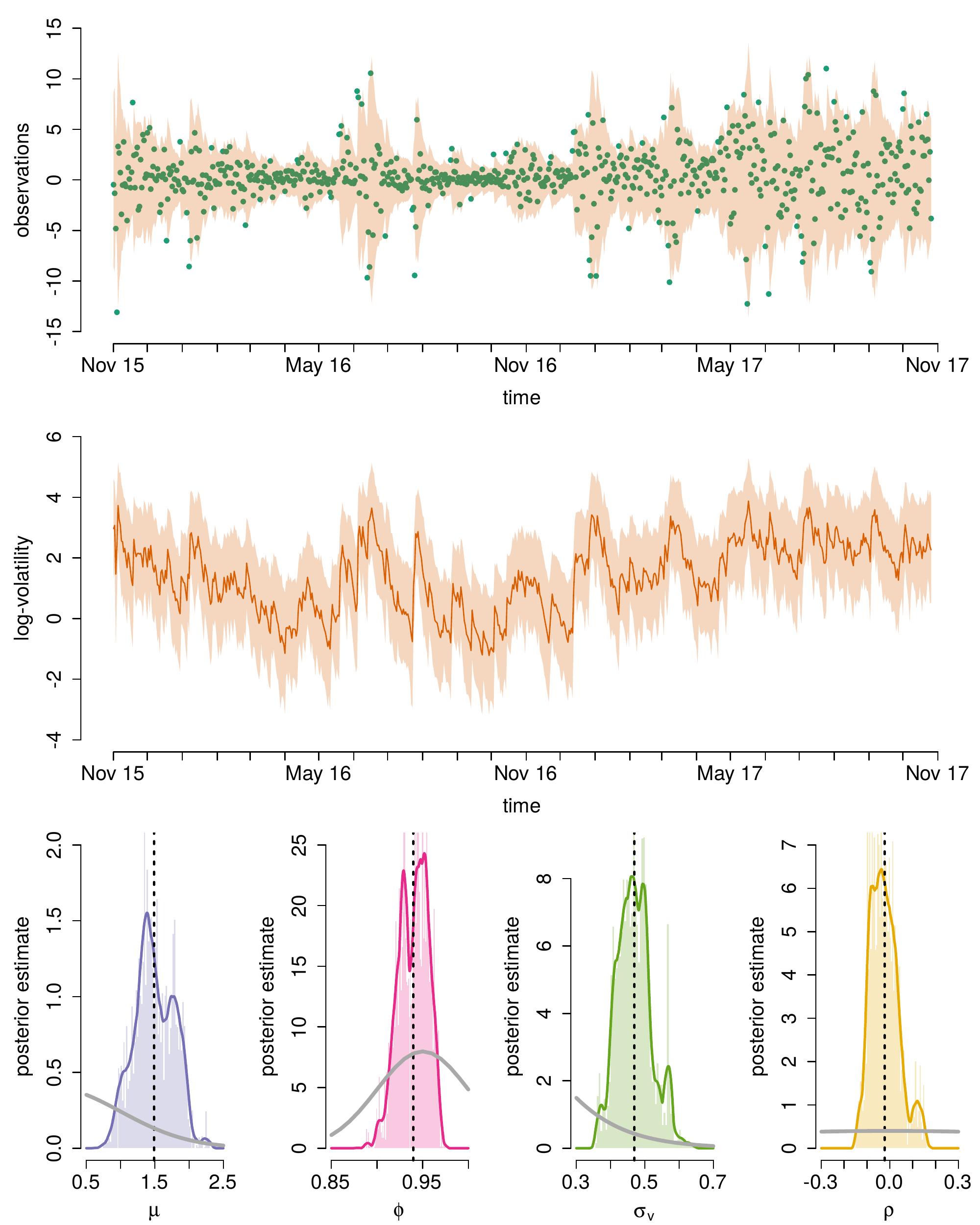}
	\caption{Top: the log-returns (green) for Bitcoin and their estimated $95\%$ confidence intervals (orange) using the model and the estimate of the log-volatility. Bottom, the posterior estimates for $\mu$ (pruple), $\phi$ (magents), $\sigma_v$ (green) and $\rho$ (yellow) obtained by dBFGS. The dotted and gray lines indicate the estimated posterior mean and the prior distributions, respectively.}
	\label{fig:example3_stochastic_volatility_paper}
\end{figure}

\subsection{SV model with Bitcoin price data}
\label{sec:results:sv}
To demonstrate a practical application of the proposed method, we consider the problem of estimating the volatility of Bitcoin prices between November 7, 2015 and November 7, 2017.
The log-returns (the change in percent of the Bitcoin price between two days) is presented in the upper part of Figure~\ref{fig:example3_stochastic_volatility_paper} as green dots.
Note that the log-returns have zero mean but that the variance is changing over time.
We aim to capture this change in volatility (as this variance is known) by the model
\begin{align*}
	\begin{bmatrix}	x_{t+1} \\ y_{t} \end{bmatrix} \Bigg| x_t
	\sim
	\mathcal{N}
	\left(
	\begin{bmatrix}	x_{t+1} \\ y_{t} \end{bmatrix};
	\begin{bmatrix}	\mu + \phi( x_t - \mu ) \\ 0 \end{bmatrix},
	\begin{bmatrix}	\sigma_v^2 & \rho \\ \rho & \exp(x_t) \end{bmatrix}
	\right),
\end{align*}
which is a so-called stochastic volatility (SV) model with leverage.
Here, the unknown latent parameters are the same as in the LGSS model together with the correlation $\rho \in (-1 ,1)$, i.e., $\theta=\{\mu,\phi,\sigma_v, \rho\}$.
The aim is to estimate the log-volatility given the data, which can be done by marginalising over the posterior estimate of the parameters, see \cite{AndrieuDoucetHolenstein2010} or \cite{DahlinSchon2017}.

The resulting estimates of the log-volaility (the latent state) is presented in the middle plot of Figure~\ref{fig:example3_stochastic_volatility_paper}.
Note that it varies over the time period and is large when the Bitcoin prices are volatile and exhibit large day-to-day changes.
Furthermore, the mean of the log-volatility process is quite large, which translates into that the log-volatility typically is large (compared with e.g., prices of stocks).
The correlation $\rho$ is probably quite close to zero.
This is quite different from $-0.8$ which is typical for stocks, which implies the large drops in stock prices raises the volatility (as investors sell their assets).

This information is very useful in many financial applications such as pricing futures on Bitcoins as well as computing various risk measures required to be presented by banks and financial institutions to regulatory agencies.


\section{Conclusions}
\label{sec:conclusions}
The numerical illustrations indicate the the proposed method can outperform many existing methods used to create good proposals for MH.
Furthermore, we would like to again underline that the proposed method requires basically no pilot runs, which are required for all pre-conditioned methods.
Moreover, the damped BFGS approach always provides a positive definite estimate of the Hessian, so no potentially numerically unstable Hessian correction is required.

Furthermore, the gradient information is crucial for estimation in large dimension parameter spaces, which are common in SSMs and in transfer function models.
All these benefits could potentially allow for a wide adoption of MH for identifying dynamical systems.

There are plenty of interesting avenues for future work within the scope of this paper.
SR1 updates \citep[Ch.\ 6.2]{NocedalWright2006} are an alternative to BFGS which are known to provide more accurate estimates of the Hessian in many cases.
Furthermore, trust region approaches from optimisation could potentially be useful in MH to protect with problems with numerical stability.
Finally, more extensive numerical evaluations are required for models with larger parameter spaces.
In this case, alternatives to or better algorithms for particle smoothing are required to obtain reasonable gradient estimates.

The source code and data used in this paper are available from GitHub \url{https://github.com/compops/qnmh-sysid2018/} and via Docker (see \texttt{README.md}).

\bibliographystyle{plainnat}
\bibliography{dahlin}

\appendix

\section{Implementation details}
\label{app:impdetails}
In Sections~\ref{sec:results:lgss:kalman} and \ref{sec:results:lgss:particles}, we use a standard Kalman filter with the RTS smoother to compute the log-posterior and its gradients.
Furthermore, a bootstrap particle filter is employed with $N=1,000$ particles and a fixed-lag smoother with lag $10$, see \cite{DahlinLindstenSchon2015a} for an algorithmic description.
The number of particles is selected using the results in \cite{DoucetPittKohn2015}.
We initialise all MH algorithms in the true parameters for simplicity and run them for $K=10,000$ iterations and discard the first $3,000$ as burn-in.

The pre-conditioning matrices for MH0 and MH1 are computed using a number of pilot runs.
The step-lengths are selected using existing rule-of-thumbs \citep{RobertsRosenthal1998} as $\epsilon_0=1.37$ and $\epsilon_1=0.57$ when using Kalman methods and $\epsilon_0=1.48$ and $\epsilon_1=0.47$ when using particle methods \citep{NemethSherlockFearnhead2016}.
For the qMH approaches, we use memory length $M=20$ and make use of a random walk proposal for the first $M$ iterations with step lengths $\epsilon=0.01$ for all three parameters.
The step size $\epsilon_2=0.5$ is used for all qMH algorithms after the initial $M$ iterations.

A reparametrization of the LGSS model is done to make all the parameters in the Markov chain unconstrained (able to assume any real value) given by
\begin{align*}
	\phi = \text{tanh}(\bar{\phi}), \qquad \sigma_v = \exp(\bar{\sigma}_v),
\end{align*}
where $\bar{\theta}=\{\mu, \bar{\phi}, \bar{\sigma}_v\}$ are the new states of the Markov chain.
This change of variables introduces a Jacobian term into the acceptance probability, see \citet[Section 6.3.2]{DahlinSchon2017}.
Finally, we use the following prior densities
\begin{align*}
	\mu \sim \mathcal{N}(0,1^2), \quad
	\phi \sim \mathcal{TN}_{(-1,1)}(0.5,1^2), \quad
	\sigma_v \sim \mathcal{G}(2,2),
\end{align*}
where $\mathcal{TN}_{(a,b)}(\cdot)$ denotes a truncated Gaussian distribution on $[a,b]$ and $\mathcal{G}(a,b)$ denotes the Gamma distribution with mean $a/b$.

In Section~\ref{sec:results:sv}, we make use of the same settings as for the LGSS model but increase $N$ to $1,500$.
Moreover, we change the priors ($p(\mu)$ is kept as before) slightly to
\begin{align*}
	\phi \sim \mathcal{TN}_{(-1,1)}(0.95,0.05), \quad
	\sigma_v \sim \mathcal{G}(2,10), \quad
	\rho \sim \mathcal{TN}_{(-1,1)}(0,1),
\end{align*}
which better reflect the parameter values usually found in real-world data.
The correlation $\rho$ is reparametrized in the same manner as $\phi$.
The Bitcoin data is computed as $y_t = 100 [ \log(s_{t}) - \log(s_{t-1}) ]$, where $s_t$ denotes the daily exchange rates versus the US Dollar obtained from \url{https://www.quandl.com/BITSTAMP/USD}.

\section{Hessian corrections}
\label{app:hessian:corrections}
The first approach \emph{regularise} the Hessian by
\begin{align*}
	\Sigma_{\text{QN}}^{\star} \big( \psi_{k, M} \big)
	=
	\Sigma_{\text{QN}} \big( \psi_{k, M} \big)
	-
	2 \lambda_{\min} I_p,
\end{align*}
with $\lambda_{\min}$ denoting the smallest (negative) eigenvalue.
This shifts all eigenvalues to be positive.
The second method \emph{flips} the eigenvalues,
\begin{align*}
	\Sigma_{\text{QN}}^{\star} \big( \psi_{k, M} \big)
	=
	Q
	|\Lambda|
	Q^{-1},
\end{align*}
where $Q$ and $|\Lambda|$ denotes the matrix of eigenvectors and the diagonal matrix of the absolute value of the eigenvalues of $\Sigma_{\text{QN}} (\psi_{k, M})$, respectively.
The third approach is the \emph{hybrid method} from \cite{DahlinLindstenSchon2015a}.
In which the estimate is replaced by a global approximation of the posterior covariance
\begin{align*}
	\Sigma_{\text{QN}}^{\star} \big( \psi_{k, M} \big)
	=
	\Sigma^{-1}_{\text{emp}},
\end{align*}
where $\Sigma_{\text{emp}}$ denotes the sample estimate of the
posterior covariance computed using the latter half of the burn-in
phase.

\end{document}